\documentstyle[11pt]{article}
\begin{document}

\title{\textbf{Generalized second law of thermodynamics in $f(T)$ gravity\thanks{K. Karami dedicated this paper
to the 80 year Jubilee of Professor Yousef Sobouti.}}}

\author{K. Karami$^{1,2}$\thanks{E-mail: KKarami@uok.ac.ir} ,
A. Abdolmaleki$^{1}$\thanks{E-mail: AAbdolmaleki@uok.ac.ir}\\\\
$^{1}$\small{Department of Physics, University of Kurdistan,
Pasdaran St., Sanandaj, Iran}\\$^{2}$\small{Center for Excellence in
Astronomy \& Astrophysics (CEAA-RIAAM ), Maragha, Iran}}

\maketitle

\begin{abstract}
We investigate the validity of the generalized second law (GSL) of
gravitational thermodynamics in the framework of $f(T)$ modified
teleparallel gravity. We consider a spatially flat FRW universe
containing only the pressureless matter. The boundary of the
universe is assumed to be enclosed by the Hubble horizon. For two
viable $f(T)$ models containing $f(T)=T+\mu_1{(-T)}^n$ and
$f(T)=T-\mu_2 T(1-e^{\beta\frac{T_0}{T}})$, we first calculate the
effective equation of state and deceleration parameters. Then, we
investigate the null and strong energy conditions and conclude that
a sudden future singularity appears in both models. Furthermore,
using a cosmographic analysis we check the viability of two models.
Finally, we examine the validity of the GSL and find that for both
models it is satisfied from the early times to the present epoch.
But in the future, the GSL is violated for the special ranges of the
torsion scalar $T$.
\end{abstract}

%\noindent\textbf{PACS numbers:} 04.50.Kd\\
\clearpage
\section{Introduction}
Observational data coming from the type Ia supernovae (SNeIa)
surveys, large scale structure (LSS), and cosmic microwave
background (CMB) anisotropy spectrum indicate that the expansion of
our present universe is accelerating rather than slowing down
\cite{Riess}. This cosmic acceleration can not be explained by the
four known fundamental interactions in the standard models, which is
the greatest challenge today in the modern physics. The proposals
that have been put forward to explain this observed phenomenon can
basically be classified into two categories. One is to assume that
in the framework of Einstein's general relativity (GR), an exotic
component with negative pressure called dark energy (DE) is
necessary to explain this observed phenomena. For a good review on
the dynamics of different DE models see \cite{Padmanabhan} and
references therein. Another alternative to account for the current
accelerating cosmic expansion is to modify GR theory. The well-known
modified gravity theories are, for examples, $f(R)$ theory,
scalar-tensor theory (including Brans-Dicke theory), braneworld
scenarios (such as DGP, RSI and RSII), $f(\mathcal{G})$ theory
($\mathcal{G}$ is the Gauss-Bonett term), Ho\v{r}ava-Lifshitz
theory, MOdified Newtonian Dynamics (MOND), and so forth. For some
relevant reviews see \cite{Tsujikawa}.

Recently, a new modified gravity theory namely the so-called $f(T)$
theory \cite{bengochea}-\cite{Li} was proposed to describe the
present accelerating expansion of the universe without resorting to
DE. It is a generalization of the teleparallel gravity (TG)
\cite{Einstein} by replacing the so-called torsion scalar $T$ with
$f(T)$. TG was originally developed by Einstein in an attempt of
unifying gravity and electromagnetism. The basic variables in TG are
tetrad fields $e_{i\mu}$, where the Weitzenbock connection rather
than the Levi-Civita connection was used to define the covariant
derivative. As a result, the spacetime has no curvature but contains
torsion. A vector $V^{\mu}$ in TG is parallel transported along a
curve if its projection $V_i=e_{i\mu}V^{\mu}$ remains constant, this
is the so-called teleparallelism. The main advantage of $f(T)$
theory is that its field equations are the second order which are
remarkably simpler than the fourth order equations of $f(R)$ theory
\cite{WufT}.

The other interesting issue in modern cosmology is the
thermodynamical description of the accelerating universe driven by
DE or modified gravity. It was shown that by applying the first law
of thermodynamics (Clausius relation) $-{\rm d}E=T_A{\rm d} S_A$ to
the apparent horizon $\tilde{r}_{\rm A}$, the Friedmann equation in
the Einstein gravity can be derived if we take the Hawking
temperature $T_{\rm A}=1/(2\pi \tilde{r}_{\rm A})$ and the entropy
$S_A=\frac{A}{4G}$ on the apparent horizon, where $A$ is the area of
the horizon \cite{Cai05}. Here, ${\rm d}E$ is the amount of energy
flow through the fixed apparent horizon. The equivalence between the
first law of thermodynamics and the Friedmann equation was also
found for gravity with Gauss-Bonnet term, the Lovelock gravity
theory and the braneworld scenarios \cite{Cai05,Akbar,Sheykhi1}.

Note that in the thermodynamics of the apparent horizon in the
Einstein gravity, the geometric entropy is assumed to be
proportional to its horizon area, $S_A=\frac{A}{4G}$ \cite{Cai05}.
However, this definition is changed for other modified gravity
theories. For instance, the geometric entropy in $f(R)$ gravity is
given by $S_A=\frac{Af_R}{4G}$ \cite{Akbar2}, where the subscript
$R$ denotes a derivative with respect to the curvature scalar R. In
$f(T)$ gravity, it was shown that when $f_{TT}$ is small, the first
law of black hole thermodynamics is satisfied approximatively and
the entropy of horizon is $S_A=\frac{Af_T}{4G}$ \cite{Miao}, where
the subscript $T$ denotes a derivative with respect to the torsion
scalar $T$.

Besides examining the validity of the thermodynamical interpretation
of gravity by expressing the gravitational field equations into the
first law of thermodynamics in different spacetimes, it is also of
great interest to investigate the validity of the generalized second
law (GSL) of thermodynamics in the accelerating universe
\cite{Izquierdo1}-\cite{Geng}. The GSL of thermodynamics like the
first law is an accepted principle in physics.

Here, our aim is to investigate the GSL of thermodynamics in the
framework of $f(T)$ gravity for a spatially flat
Friedmann-Robertson-Walker (FRW) universe filled with the
pressureless matter. To do this, in section 2, we briefly review the
$f(T)$ gravity. In section 3, we investigate the GSL of
thermodynamics on the dynamical apparent horizon with the Hawking
temperature. In section 4, for two viable $f(T)$ models we first
calculate the effective equation of state and deceleration
parameters. Then, the null and strong energy conditions are
investigated. Also the viability of both models is checked by
cosmography. Finally, the validity of the GSL is examined. Section 5
is devoted to conclusions.

%--------------------------------------------------------------------------------------------------------------------
\section{Brief review of $f(T)$ gravity}
In the framework of $f(T)$ theory, the action of modified TG is
given by \cite{Ferraro}
\begin{equation}
I =\frac{1}{16\pi G}\int {\rm d}^4
x~e~\Big[f(T)+L_m\Big],\label{action}
\end{equation}
where $e={\rm det}(e^i_{\mu})=\sqrt{-g}$ and $L_m$ is the Lagrangian
density of the matter inside the universe. Also $e^i_{\mu}$ is the
vierbein field which is used as a dynamical object in TG.

The modified Friedmann equations in the framework of $f(T)$ gravity
in the spatially flat FRW universe are given by \cite{WufT,KA}
\begin{equation}
H^2=\frac{8\pi G}{3}(\rho_m+\rho_T),\label{fT11}
\end{equation}
\begin{equation}
\dot{H}+\frac{3}{2}H^2=-4\pi G(p_m+p_T),\label{fT22}
\end{equation}
where
\begin{equation}
\rho_T=\frac{1}{16\pi G}(2Tf_T-f-T),\label{roT}
\end{equation}
\begin{equation}
p_T=-\frac{1}{16\pi
G}[-8\dot{H}Tf_{TT}+(2T-4\dot{H})f_T-f+4\dot{H}-T],\label{pT}
\end{equation}
\begin{equation}
T=-6H^2,\label{T}
\end{equation}
and $H=\dot{a}/a$ is the Hubble parameter. Here $\rho_m$ and $p_m$
are the energy density and pressure of the matter, respectively.
Also $\rho_T$ and $p_T$ are the torsion contributions to the energy
density and pressure. The energy conservation laws are still given
by
\begin{equation}\label{CEqm}
\dot{\rho}_m+3H(\rho_m+p_m)=0,
\end{equation}
\begin{equation}
\dot{\rho}_T+3H(\rho_T+p_T)=0.\label{ecT}
\end{equation}
Note that if $f(T)=T$, from Eqs. (\ref{roT}) and (\ref{pT}) we have
$\rho_T=0$ and $p_T=0$ then Eqs. (\ref{fT11}) and (\ref{fT22})
transform to the usual Friedmann equations in GR.

The effective equation of state (EoS) parameter due to the torsion
contribution is defined as \cite{WufT,KA}
\begin{equation}
\omega_T=\frac{p_T}{\rho_T}=-1-\frac{\dot{T}}{3H}\left(\frac{2Tf_{TT}+f_T-1}{2Tf_T-f-T}\right).\label{omegaT}
\end{equation}
Note that for the de Sitter universe, i.e. $\dot{H}=0=\dot{T}$, we
have $\omega_T=-1$ behaving like the cosmological constant.

Using Eqs. (\ref{fT11}), (\ref{roT}) and (\ref{T}) we have
\begin{equation}
\rho_m=\frac{1}{16\pi G}(f-2Tf_T).\label{rhom}
\end{equation}
Here, we consider a spatially flat FRW universe filled with the
pressureless matter, i.e. $p_m=0$. We ignore the contribution of
radiation which is compatible with the observations \cite{Egan}.
From Eqs. (\ref{fT11}) to (\ref{pT}) for $p_m=0$, one can obtain
\begin{equation}
\dot{H}=-\frac{4\pi G\rho_m}{f_T+2Tf_{TT}}.\label{Hdot}
\end{equation}
Substituting Eq. (\ref{rhom}) into (\ref{Hdot}) and using
$\dot{T}=-12H\dot{H}$, one can get
\begin{equation}
\dot{T}=3H\left(\frac{f-2Tf_{T}}{f_T+2Tf_{TT}}\right).\label{Tdot}
\end{equation}
With the help of above relation, the effective EoS parameter
(\ref{omegaT}) can be rewritten as
\begin{equation}
\omega_T=-\frac{f/T-f_T+2Tf_{TT}}{(f_T+2Tf_{TT})(f/T-2f_T+1)}.\label{omegaT2}
\end{equation}
For the deceleration parameter
\begin{equation}
q=-1-\frac{\dot{H}}{H^2},\label{q1}
\end{equation}
using Eqs. (\ref{T}) and (\ref{Tdot}) one can obtain
\begin{equation}
q=2\left(\frac{f_T-Tf_{TT}-\frac{3f}{4T}}{f_T+2Tf_{TT}}\right).\label{q2}
\end{equation}
For the case $f(T)=T$, we get $q=0.5$ corresponding to the matter
dominated phase.
%--------------------------------------------------------------------------------------------------------------------
\section{GSL in $f(T)$ gravity}
Here, we investigate the validity of the GSL for a spatially flat
FRW universe filled with the pressureless barionc matter (BM) and
dark matter (DM). According to the GSL, the entropy of matter inside
the horizon plus the entropy of the horizon must not decrease in
time \cite{Cai05}. We assume that the boundary of the universe to be
enclosed by the dynamical apparent horizon $\tilde{r}_{\rm A}$. Also
the Hawking temperature on the apparent horizon $\tilde{r}_{\rm A}$
is given by \cite{Cai05}
\begin{equation}
T_{\rm A}=\frac{1}{2\pi \tilde{r}_{\rm
A}}\left(1-\frac{\dot{\tilde{r}}_{\rm A}}{2H\tilde{r}_{\rm A}}
\right),\label{TA1}
\end{equation}
where $\frac{\dot{\tilde{r}}_{\rm A}}{2H\tilde{r}_{\rm A}}<1$
ensures that the temperature is positive.

The entropy of the universe including BM and DM inside the dynamical
apparent horizon is given by Gibb's equation \cite{Izquierdo1}
\begin{equation}
T_{\rm A}{\rm d}S_{m}={\rm d}E_{m}+p_{m}{\rm d}V,\label{Tdsm1}
\end{equation}
where $p_{m}=0$ and ${\rm V}=4\pi \tilde{r}_{\rm A}^3/3$ is the
volume containing the pressureless matter with the radius of the
dynamical apparent horizon $\tilde{r}_{\rm A}$. Also
\begin{equation}
E_{m}=\frac{4\pi \tilde{r}_{\rm A}^3}{3}\rho_{m}.\label{Em}
\end{equation}
Taking time derivative of both sides of Eq. (\ref{Em}) and using
(\ref{CEqm}), Gibb's equation (\ref{Tdsm1}) yields
\begin{equation}
T_{\rm A}\dot{S}_{m}=4\pi{\tilde{r}_{\rm
A}^2}\rho_{m}({\dot{\tilde{r}}_{\rm A}}-H\tilde{r}_{\rm
A}),\label{Tdsm2}
\end{equation}
where the dot denotes a time derivative. Substituting Eq.
(\ref{rhom}) into (\ref{Tdsm2}) gives
\begin{equation}
T_{\rm A}\dot{S}_{m}=\frac{\tilde{r}_{\rm
A}^2}{4G}({\dot{\tilde{r}}_{\rm A}}-H\tilde{r}_{\rm
A})(f-2Tf_T).\label{Tdsm3}
\end{equation}
Now we need to calculate the contribution of the apparent horizon
entropy. Following \cite{Miao}, when $f_{TT}$ is small, the horizon
entropy in $f(T)$ gravity is given by
\begin{equation}
S_A=\frac{Af_T}{4G},\label{SA}
\end{equation}
where $A=4\pi \tilde{r}_{\rm A}^2$. For the special case $f(T)=T$,
Eq. (\ref{SA}) recovers the Bekenstein-Hawking entropy
$S_A=\frac{A}{4G}$ in the Einstein gravity.

Using Eq. (\ref{TA1}), the evolution of horizon entropy (\ref{SA})
is obtained as
\begin{equation}
T_{\rm A}\dot{S}_{\rm
A}=\frac{1}{2G}\left(1-\frac{\dot{\tilde{r}}_{\rm
A}}{2H\tilde{r}_{\rm A}}\right)\Big(2\dot{\tilde{r}}_{\rm
A}f_T+\tilde{r}_{\rm A}\dot{T}f_{TT}\Big).\label{TAdSA1}
\end{equation}
Inserting Eq. (\ref{Tdot}) into (\ref{TAdSA1}) gives
\begin{equation}
T_{\rm A}\dot{S}_{\rm
A}=\frac{1}{2G}\left(1-\frac{\dot{\tilde{r}}_{\rm
A}}{2H\tilde{r}_{\rm A}}\right)\left[\frac{(2\dot{\tilde{r}}_{\rm
A}-3H\tilde{r}_{\rm A})2Tf_Tf_{TT}+3H\tilde{r}_{\rm
A}f(T)f_{TT}+2\dot{\tilde{r}}_{\rm
A}f_T^2}{f_T+2Tf_{TT}}\right].\label{TAdSA2}
\end{equation}
For the flat FRW metric, the dynamical apparent horizon is same as
the Hubble horizon given by \cite{Poisson}
\begin{equation}
\tilde{r}_{\rm A}=\frac{1}{H}.\label{rA}
\end{equation}
Taking time derivative of Eq. (\ref{rA}), using (\ref{T}),
(\ref{rhom}) and (\ref{Hdot}) one can get
\begin{equation}
\dot{\tilde{r}}_{\rm
A}=-\frac{3}{2T}\left(\frac{f-2Tf_T}{f_T+2Tf_{TT}}\right).\label{rdotA}
\end{equation}
Replacing Eqs. (\ref{rA})-(\ref{rdotA}) into (\ref{Tdsm3}) and
(\ref{TAdSA2}) and using (\ref{T}), one can obtain
\begin{equation}
T_{\rm
A}\dot{S}_{m}=\frac{3}{4GT^2}\left(\frac{f-2Tf_T}{f_T+2Tf_{TT}}\right)\Big(4T^2f_{TT}-4Tf_T+3f\Big),\label{Tdsm4}
\end{equation}
\begin{equation}
T_{\rm A}\dot{S}_{\rm
A}=\frac{3}{4GT^2}\left(\frac{f-2Tf_T}{f_T+2Tf_{TT}}\right)
\left[\frac{\Big(4T^2f_{TT}^2-Tf_T+\frac{3}{2}f\Big)\Big(2T^2f_{TT}^2-f_T^2-Tf_Tf_{TT}\Big)}{(f_T+2Tf_{TT})^2}\right]
.\label{TAdSA3}
\end{equation}
Adding Eqs. (\ref{Tdsm4}) and (\ref{TAdSA3}) yields the GSL in
$f(T)$ gravity as
\begin{equation}
T_{\rm A}\dot{S}_{\rm
tot}=\frac{9}{8G}\left(\frac{f-2Tf_T}{f_T+2Tf_{TT}}\right)\left[
4f_{TT}+\left(\frac{f-2Tf_T}{f_T+2Tf_{TT}}\right)\left(\frac{f_T+5Tf_{TT}}{T^2}\right)\right]
,\label{TASdot}
\end{equation}
where $S_{\rm tot}=S_m+S_{\rm A}$ is the total entropy due to
different contributions of the matter and the horizon. Note that in
the Einstein TG, i.e. $f(T)=T$, the GSL (\ref{TASdot}) yields
\begin{equation}
T_{\rm A}\dot{S}_{\rm tot}=\frac{9}{8G}>0,
\end{equation}
which always holds. In what follows, we investigate the validity of
the GSL, i.e. $T_{\rm A}\dot{S}_{\rm tot}\geq 0$, for two viable
$f(T)$ models introduced in the literature.
%--------------------------------------------------------------------------------------------------------------------
\section{Two viable $f(T)$ models}
Here, we consider two viable $f(T)$ models proposed by
\cite{WufT1-2,linder} to explain the present cosmic accelerating
expansion of the universe. The first model is a power law
\begin{equation}
f(T)=T+\mu_1{(-T)}^n,~~~~~~{\rm Model~1},\label{model1}
\end{equation}
where $\mu_1$ and $n$ are constants \cite{WufT1-2,linder}. The
second model has an exponential dependence on the torsion scalar as
\begin{equation}
f(T)=T-\mu_2 T\Big(1-e^{\beta\frac{T_0}{T}}\Big),~~~~~~{\rm
Model~2},\label{model2}
\end{equation}
where $\mu_2$ and $\beta$ are two model parameters, and
$T_0=-6H_0^2$ \cite{WufT1-2}.

The parameters $\mu_1$ and $\mu_2$ can be obtained by inserting Eqs.
(\ref{model1}) and (\ref{model2}) into the modified Friedmann Eq.
(\ref{fT11}). Solving the resulting equations for the present time
gives
\begin{equation}
\mu_1=\left(\frac{1-\Omega_{m_0}}{2n-1}\right)(6H_0^2)^{1-n}
,\label{mu1}
\end{equation}
\begin{equation}
\mu_2=\frac{1-\Omega_{m_0}}{1-(1-2\beta)e^\beta},\label{mu2}
\end{equation}
where $\Omega_{m_0}=\frac{8\pi G\rho_{m_0}}{3H_0^2}$ is the
dimensionless matter energy density and the index 0 denotes the
value of a quantity at the present. Note that both models can unify
a number of interesting extensions of gravity beyond the standard
GR. For instance, model 1 for the cases $n=0$ and $n=1/2$ reduces to
the $\Lambda$CDM and DGP \cite{Dvali} models, respectively. Model 2
acts like the $\Lambda$CDM for $\beta=0$. Also both models can
satisfy the condition
\begin{equation}
\lim_{T\rightarrow\infty}f/T\rightarrow 1,
\end{equation}
at high redshift which is compatible with the primordial
nucleosynthesis and CMB constraints \cite{WufT1-2,linder}. For model
1 to be a viable model compared to current data one needs $n\ll 1$
\cite{WufT1-2,linder}. The joint analysis of the astronomical data
from SNeIa+BAO+CMB gives the best fit values
($n=0.04_{-0.33}^{+0.22}, \Omega_{m_0}=0.272_{-0.032}^{+0.036}$) for
model 1 and ($\beta=-0.02_{-0.20}^{+0.31},
\Omega_{m_0}=0.272_{-0.034}^{+0.036}$) for model 2 at the 95\%
confidence level (CL) \cite{WufT1-2}.

Using the statefinder geometrical analysis and $Om(z)$ diagnostic
method, it was shown that both model 1 and model 2 evolve from the
standard cold DM (SCDM) to a de Sitter phase \cite{WufT1-2}. Also
the effective DE for model 2 with $\beta\neq 0$ is similar to the
$\Lambda$CDM both in the high redshift regimes and in the future,
while for model 1 with $n\neq0$ this similarity occurs only in the
future \cite{WufT1-2}.

\begin{figure}
\includegraphics{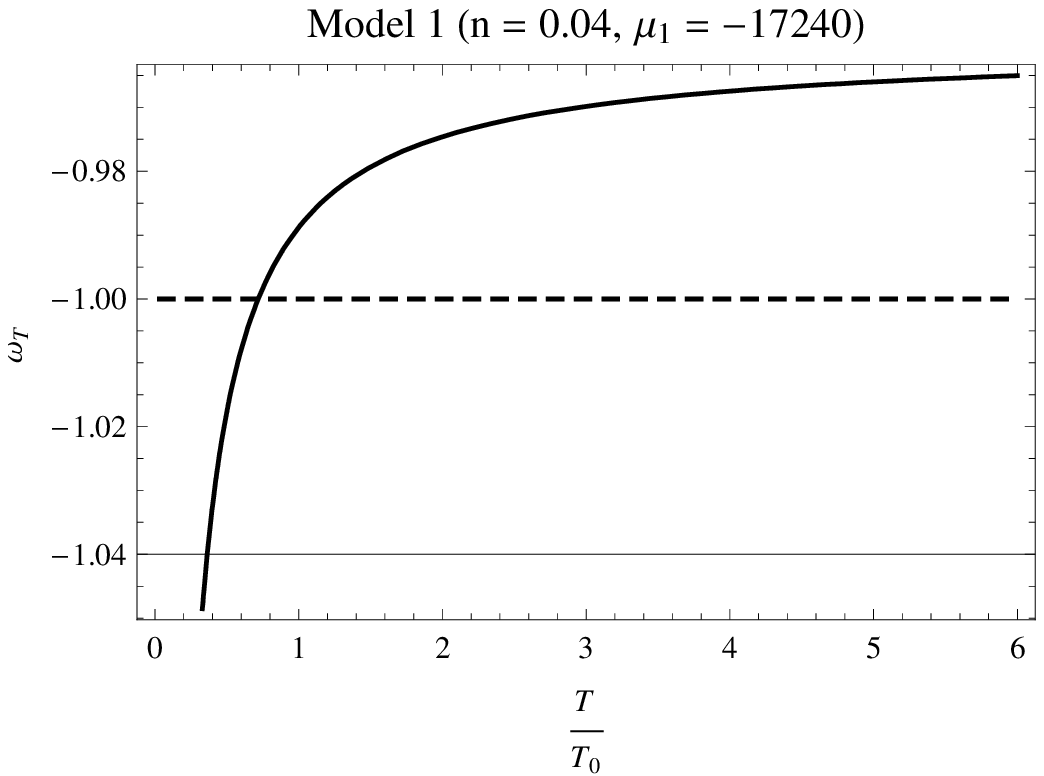}
      \vspace{7.0cm}
\caption[]{The evolution of the effective EoS parameter, Eq.
(\ref{omegaT2}), versus $\frac{T}{T_0}$ for model 1, Eq.
(\ref{model1}). Auxiliary parameters are: $n=0.04$,
$\Omega_{m_0}=0.272$ \cite{WufT1-2}, $H_0=74.2~{\rm
Km~S^{-1}~Mpc^{-1}}$ \cite{Riess99}. For these values one finds
$\mu_1=\left(\frac{1-\Omega_{m_0}}{2n-1}\right)(6H_0^2)^{1-n}=-17240$.}
         \label{wT-model1}
   \end{figure}

\begin{figure}
\includegraphics{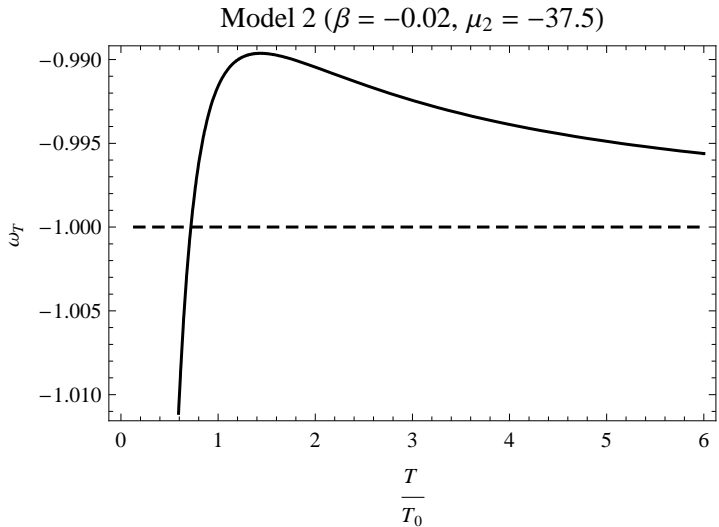}
      \vspace{7.0cm}
\caption[]{Same as Fig. \ref{wT-model1}, for model 2, Eq.
(\ref{model2}). Auxiliary parameters are: $\beta=-0.02$,
$\Omega_{m_0}=0.272$ \cite{WufT1-2}, $H_0=74.2~{\rm
Km~S^{-1}~Mpc^{-1}}$ \cite{Riess99}. This values gives
$\mu_2=\frac{1-\Omega_{m_0}}{1-(1-2\beta)e^\beta}=-37.5$.}
         \label{wT-model2}
   \end{figure}

The evolution of the effective EoS parameter, Eq. (\ref{omegaT2}),
for model 1 and model 2 is plotted in Figs. \ref{wT-model1} and
\ref{wT-model2}, respectively. Figures show that from the early
times to the present epoch, both models behave like quintessence DE,
i.e. $\omega_T>-1$, and their effective EoS parameters can not cross
the phantom divide line \cite{Cai:2009zp}. This result has been
already obtained by \cite{WufT1-2,linder}. At early times
($T/T_0\rightarrow+\infty$), for model 1 and model 2 we have
$\omega_T\simeq-0.96$ and $-1$, respectively. Model 2 behaves like
the $\Lambda$CDM at high redshift. At the present time ($T/T_0=1$),
for model 1 and model 2 we obtain $\omega_{T_0}\simeq-0.989$ and
$-0.992$, respectively. An interesting result which is absent in
\cite{WufT1-2,linder} is that in the future ($1<T/T_0<0$) for model
1 and model 2 at $T/T_0\simeq0.718$ and $0.719$, respectively, we
have a transition from the quintessence state, $\omega_T>-1$, to the
phantom regime, $\omega_T<-1$.

\begin{figure}
\includegraphics{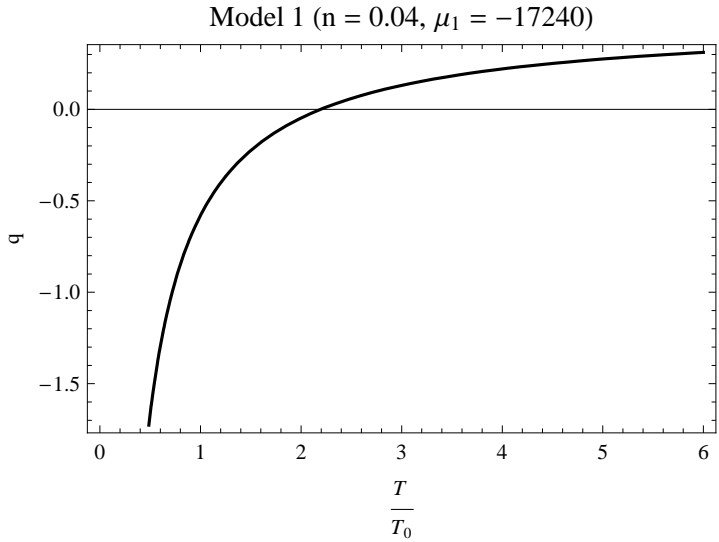}
      \vspace{7.0cm}
\caption[]{The evolution of the deceleration parameter, Eq.
(\ref{q2}), versus $\frac{T}{T_0}$ for model 1. Auxiliary parameters
as in Fig. \ref{wT-model1}.}
         \label{q-model1}
   \end{figure}

\begin{figure}
\includegraphics{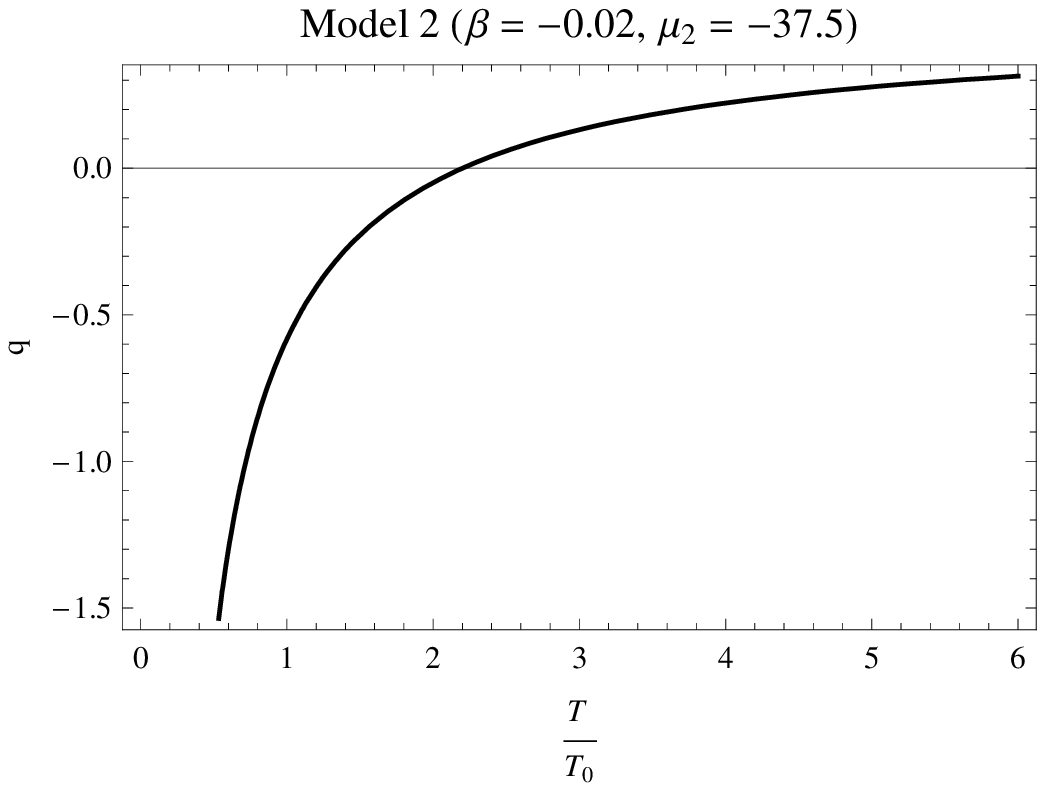}
      \vspace{7.0cm}
\caption[]{Same as Fig. \ref{q-model1}, for model 2. Auxiliary
parameters as in Fig. \ref{wT-model2}.}
         \label{q-model2}
   \end{figure}

The evolution of the deceleration parameter, Eq. (\ref{q2}), for
model 1 and model 2 is plotted in Figs. \ref{q-model1} and
\ref{q-model2}, respectively. Figures illustrate that for model 1
and model 2 in the past $T/T_0\simeq2.194$ and $2.199$,
respectively, we have a cosmic deceleration $q>0$ to acceleration
$q<0$ transition which is compatible with the observations
\cite{Ishida}. At early times ($T/T_0\rightarrow+\infty$), for both
models we obtain $q\simeq 0.5$ which indicates that the universe was
experiencing a phase of deceleration at the early stage of its
evolution due to the domination of the matter component. At the
present time ($T/T_0=1$), for model 1 and model 2 we get
$q_0\simeq-0.580$ and $-0.583$, respectively, which are in good
agreement with the recent observational result $-1.4\leq q_0\leq
-0.3$ \cite{Ishida}.

\begin{figure}
\includegraphics{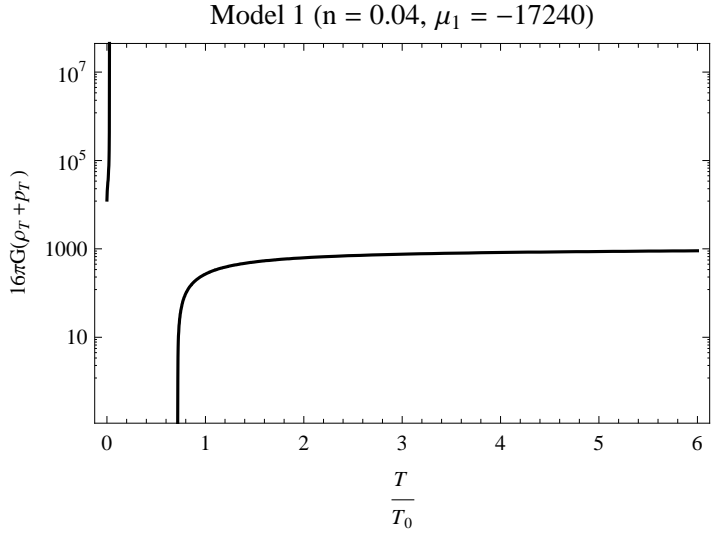}
      \vspace{7.0cm}
\caption[]{The evolution of $\rho_T+p_T$ versus $\frac{T}{T_0}$ for
model 1. Auxiliary parameters as in Fig. \ref{wT-model1}.}
         \label{NEC-model1}
   \end{figure}

\begin{figure}
\includegraphics{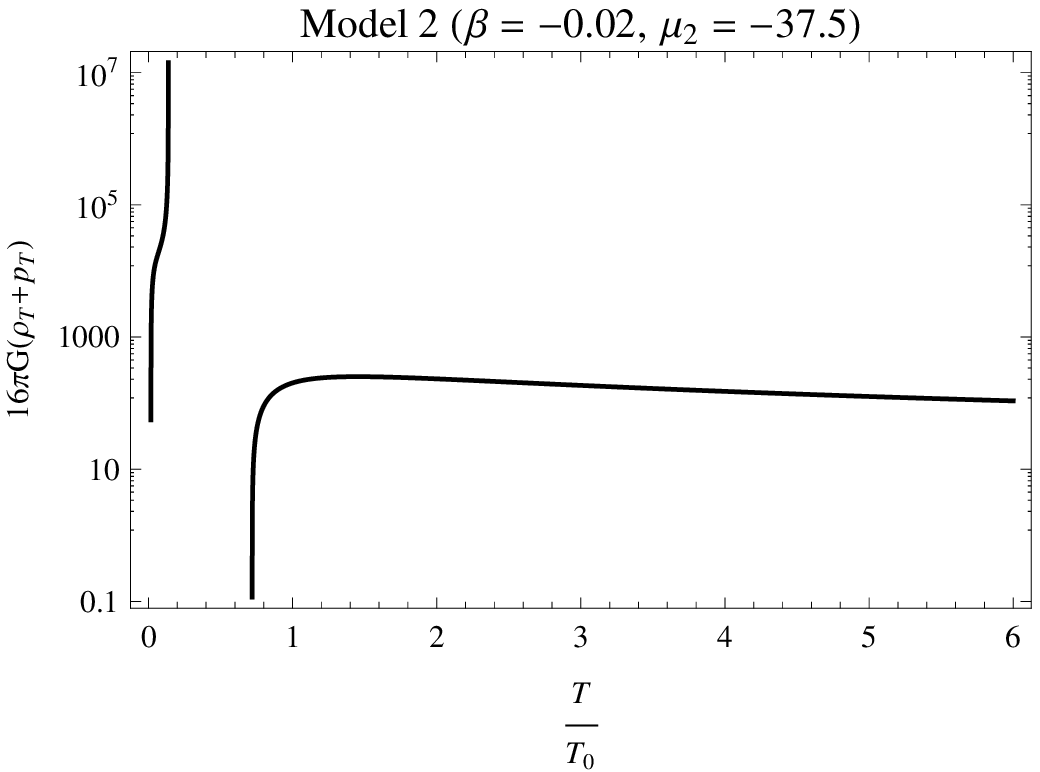}
      \vspace{7.0cm}
\caption[]{Same as Fig. \ref{NEC-model1}, for model 2. Auxiliary
parameters as in Fig. \ref{wT-model2}.}
         \label{NEC-model2}
   \end{figure}

The evolution of $\rho_T+p_T$ versus $\frac{T}{T_0}$ for model 1 and
model 2 is plotted in Figs. \ref{NEC-model1} and \ref{NEC-model2},
respectively. Figures show that the null energy condition (NEC),
i.e. $\rho_T+p_T\geq 0$, is violated for model 1 when
$T/T_0\in(0.025,0.718)$ and for model 2 when $T/T_0\in(0,0.017)$ and
$(0.138,0.719)$. When the EoS parameter of DE is less than $-1$, the
universe reaches a Big Rip singularity within a finite time. In this
case, the NEC is violated. Barrow \cite{Barrow} showed that a
different type of future singularity, the so-called sudden
singularity, can appear at a finite time even when the strong energy
condition (SEC), i.e. $\rho_T+3p_T\geq 0$ and $\rho_T+p_T\geq 0$, is
satisfied. This type of future singularity corresponds to the one in
which the pressure density $p_T$ diverges at $T=T_s$, but the energy
density $\rho_T$ is finite. The properties of different future
singularities in the DE universe have been investigated in ample
detail by \cite{Nojiri}. Our numerical results show that the sudden
future singularity appears in model 1 and model 2 when
$(T_s/T_0=0.025,16\pi G\rho_T=20754)$ and $(T_s/T_0=0.138,16\pi
G\rho_T=19822)$, respectively.  Also the SEC for model 1 and model 2
is satisfied only when $T/T_0\in(0,0.025)$ and
$T/T_0\in(0.018,0.138)$, respectively.
%--------------------------------------------------------------------------------------------------------------------
\subsection{Cosmographic analysis}
Here, following Capozziello et al. \cite{CapCosmo} we use a
cosmographic analysis to check the viability of model 1 and model 2
without the need of explicitly solving the field equations and
fitting the data. In this approach, the parameters of a given $f(T)$
model must be chosen in such a way that the model-independent
constraints on the cosmographic parameters ($h,q_0,j_0,s_0,l_0$)
obtained by fitting to SNeIa Hubble diagram and BAO data are
satisfied. Here, $h$ is the Hubble constant and $q_0,j_0,s_0,l_0$
are the deceleration, jerk, snap, and lerk parameters, respectively.
Let us first start with model 2. Using Eqs. (4.21) and (4.22) in
\cite{CapCosmo} one can find
\begin{equation}
\mu_2=\left(\frac{1}{\beta}-1\right)(1-\Omega_{m_0}).
\end{equation}
Now the $f_i=f^{(i)}(T_0)/(6H_0^2)^{-(i-1)}$ value for $i=(2,3,4,5)$
where $f^{(i)}(T)=d^if/dT^i$ can be expressed as function of $\beta$
only when we fix $\Omega_{m_0}=0.1329/h^2$ from the WMAP7 data.
Following \cite{CapCosmo} for each $f_2$ value of the sample
obtained above from the cosmographic parameters analysis, we solve
$\hat{f}_2(\beta)=f_2$. This yields $\beta=-0.2$ which takes place
in the 95\% CL from the model-dependent constraints \cite{WufT1-2}.
Then, we estimate the theoretically expected values for the other
derivatives $(f_3,f_4,f_5)$. The median and 68\% and 95\% confidence
ranges are obtained as
\begin{eqnarray}
&&f_3=0.371_{-0.351-0.914}^{+0.117+0.148}\nonumber\\
&&f_4=1.385_{-1.302-3.709}^{+0.380+0.473}\nonumber\\
&&f_5=6.447_{-6.033-18.852}^{+1.498+1.832}.
\end{eqnarray}
Now we compare the above results with the model-independent
constraints on the $f_i$ values given in Table II in
\cite{CapCosmo}. Following \cite{CapCosmo} we use only the 68\% CL
which we compare the above constraints to. Our comparison shows that
the values of $(f_3,f_4,f_5)$ take place in the 68\% CL in Table II
in \cite{CapCosmo}. Therefore, we conclude that model 2 is favored
by the observational data.

For model 1, its cosmographic analysis has been already done by
Capozziello et al. \cite{CapCosmo}. They showed that for the best
fit value of the cosmographic parameters, solving $\hat{f}_2(n)=f_2$
yields $n=-0.011$. Note that the value of $n$ takes place in the
95\% CL from the model-dependent constraints \cite{WufT1-2}.
Capozziello et al. \cite{CapCosmo} found quite small values for
$(f_3,f_4,f_5)$ as expected for the $\Lambda$CDM model. Here, the
disagreement with the constraints in Table II in \cite{CapCosmo} may
be due to this fact that depending on the value of $n$, assuming
$f^{(i)}(T_0)=0$ for $i\geq 6$ in Taylor expanding $f(T)$ can fail
for model 1 so that the constraints on $f_i$ should not be
considered reliable \cite{CapCosmo}.
%--------------------------------------------------------------------------------------------------------------------
\subsection{Examining the GSL for model 1 and model 2}
\begin{figure}
\includegraphics{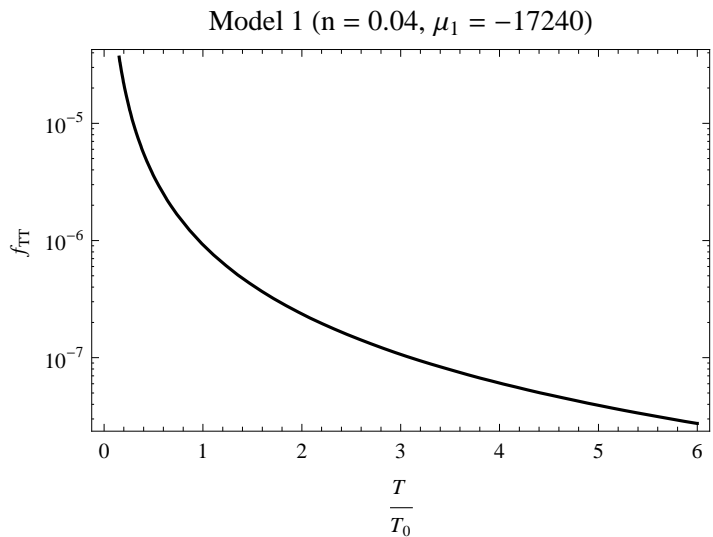}
      \vspace{7.0cm}
\caption[]{$f_{TT}$ versus $\frac{T}{T_0}$ for model 1. Auxiliary
parameters as in Fig. \ref{wT-model1}.}
         \label{fTTf-model1}
   \end{figure}

\begin{figure}
\includegraphics{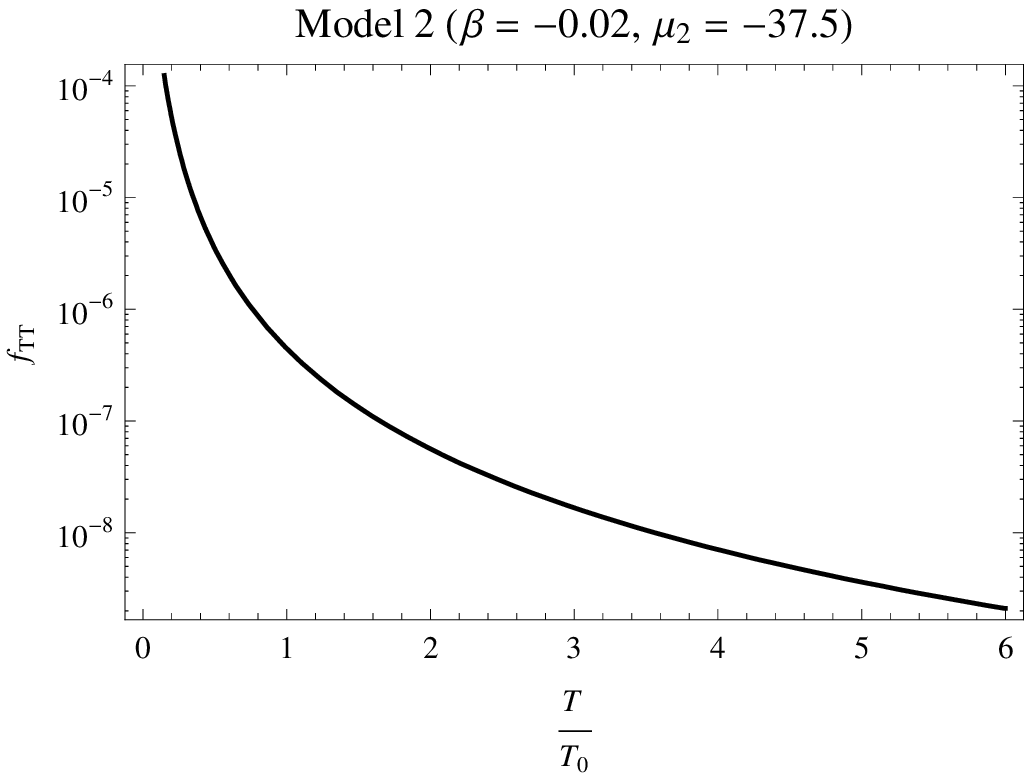}
      \vspace{7.0cm}
\caption[]{Same as Fig. \ref{fTTf-model1}, for model 2. Auxiliary
parameters as in Fig. \ref{wT-model2}.}
         \label{fTTf-model2}
   \end{figure}

Here, we examine the validity of the GSL, Eq. (\ref{TASdot}), for
both models. First, we need to check the validity of the horizon
entropy relation (\ref{SA}). Because as we already mentioned, Eq.
(\ref{SA}) is valid only when $f_{TT}$ is small \cite{Miao}. To
check this, we plot $f_{TT}$ versus $T/T_0$ for model 1 and model 2
in Figs. \ref{fTTf-model1} and \ref{fTTf-model2}, respectively.
Figures show that the $f_{TT}$ is very small throughout history of
the universe. This confirms the validity of Eq. (\ref{SA}) for both
models.

\begin{figure}
\includegraphics{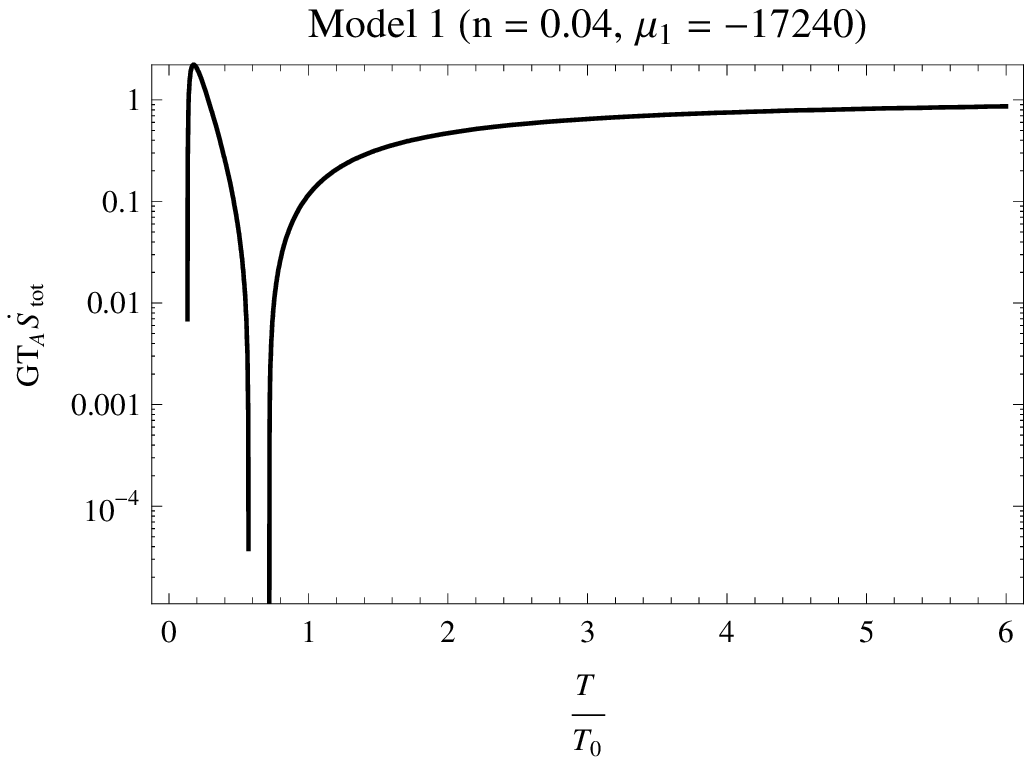}
      \vspace{7.0cm}
\caption[]{The evolution of the GSL, Eq. (\ref{TSdotmodel1}), versus
$\frac{T}{T_0}$ for model 1. Auxiliary parameters as in Fig.
\ref{wT-model1}.}
         \label{GSL-model1}
   \end{figure}

\begin{figure}
\includegraphics{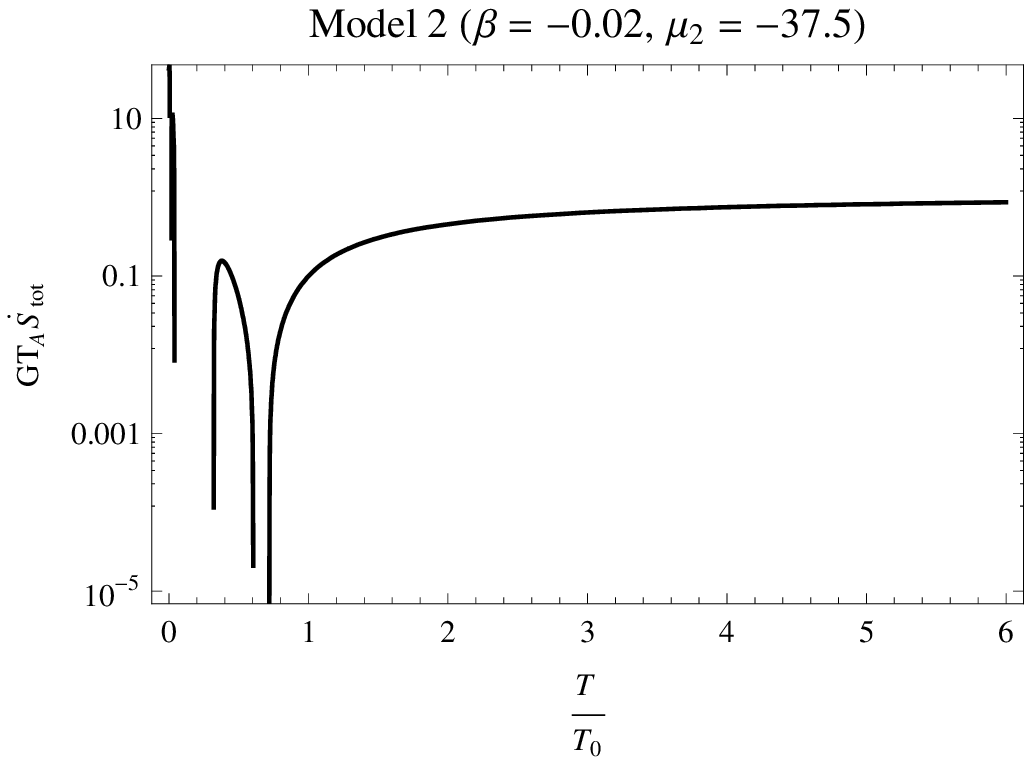}
      \vspace{7.0cm}
\caption[]{The evolution of the GSL, Eq. (\ref{TSdotmodel2}), versus
$\frac{T}{T_0}$ for model 2. Auxiliary parameters as in Fig.
\ref{wT-model2}.}
         \label{GSL-model2}
   \end{figure}

Now we can calculate the GSL, Eq. (\ref{TASdot}), for both models.
For model 1, the resulting GSL is
\begin{eqnarray}
T_{\rm A}\dot{S}_{\rm
tot}&=&\frac{9}{8}\left[\frac{1+\mu_1(1-2n){(-T)}^{n-1}}{\Big(1+\mu_1
n(1-2n){(-T)}^{n-1}\Big)^2}\right]\nonumber\\&&\times\Big[1+
\Big(1-n(2+n)\Big)\mu_1{(-T)}^{n-1}+n(1-2n)\Big(4-n(9-4n)\Big)\mu_1^2{(-T)}^{2n-2}\Big]
,\label{TSdotmodel1}
\end{eqnarray}
and for model 2 we obtain
\begin{eqnarray}
T_{\rm A}\dot{S}_{\rm tot}=\frac{\rm
I}{8T{\left[(1-\mu_2)T^2+\mu_2(T^2-\beta
T_0T+2\beta^2T_0^2)e^{\frac{\beta
T_0}{T}}\right]}^2},\label{TSdotmodel2}
\end{eqnarray}
where
\begin{eqnarray}
{\rm I}&=&9(1-\mu_2)^3T^5+9\mu_2\Big[(3-2\mu_2)T^5-5\beta
T_0(1+\mu_2-\mu_2^2)T^4
\nonumber\\&&+\beta^2T_0^2(1+17\mu_2-9\mu_2^2)T^3
-4\mu_2\beta^3T_0^3\Big((5-2\mu_2)T \nonumber\\&&+\beta
T_0(1-2\mu_2)\Big)T+16\mu_2^2T_0^5\beta^5\Big]e^{\frac{\beta
T_0}{T}}.
\end{eqnarray}
The evolutions of the GSL (\ref{TSdotmodel1}) and
(\ref{TSdotmodel2}) are plotted in Figs. \ref{GSL-model1} and
\ref{GSL-model2}, respectively. Figures clear that for both model 1
and model 2, the GSL is satisfied from the early times to the
present epoch. At early times ($T/T_0\rightarrow+\infty$), for both
models we have $GT_{\rm A}\dot{S}_{\rm tot}\simeq$ 1.125. At the
present time ($T/T_0=1$), for model 1 and model 2 we get $GT_{\rm
A}\dot{S}_{\rm tot}\simeq$ 0.116 and 0.010, respectively. Note that
in the future ($1<T/T_0<0$), Fig. \ref{GSL-model1} clears that the
GSL for model 1 is violated, i.e. $GT_{\rm A}\dot{S}_{\rm tot}<0$,
for $T/T_0\in(0,0.133)$ and $(0.569,0.718)$. Also in the future,
according to Fig. \ref{GSL-model2} the GSL for model 2, like model
1, is violated for $T/T_0\in(0.005,0.017)$, $(0.039,0.321)$ and
$(0.603,0.719)$. Figures \ref{GSL-model1} and \ref{GSL-model2} don't
show the negative values of the GSL, because the axis of $GT_{\rm
A}\dot{S}_{\rm tot}$ is based on a logarithmic scale.
%--------------------------------------------------------------------------------------------------------------------
\section{Conclusions}
Here, we studied the GSL in the framework of $f(T)$ gravity. Among
other approaches related with a variety of
 DE models, a very promising approach to DE is related
with the modified TG known as $f(T)$ gravity, in which DE emerges
from the modification of torsion. The class of $f(T)$ gravity
theories is an intriguing generalization of Einstein's new GR,
taking a curvature-free approach and using a connection with
torsion. It is analogous to the $f(R)$ extension of the
Einstein-Hilbert action of standard GR, but has the advantage of the
second order field equations \cite{linder}. We investigated the GSL
on the Hubble horizon with the Hawking temperature for a spatially
flat FRW universe filled with the pressureless matter. For two
viable $f(T)$ models containing $f(T)=T+\mu_1{(-T)}^n$ and
$f(T)=T-\mu_2 T(1-e^{\beta\frac{T_0}{T}})$, we first calculated the
effective EoS and deceleration parameters. Interestingly enough, we
found that for both models there is a transition from the
quintessence state, $\omega_T>-1$, to the phantom regime,
$\omega_T<-1$, in the future. Also both models showed a cosmic
deceleration $q>0$ to acceleration $q<0$ transition in the near past
which is compatible with the observations \cite{Ishida}.
Furthermore, we investigated the NEC and SEC and concluded that both
models show a sudden future singularity. Also our cosmographic
analysis cleared that model 1 is unable to predict the
observationally motivated $(f_2,f_3,f_4,f_5)$ values but model 2 is
favored by the observational data. Finally, we examined the validity
of the GSL for the selected $f(T)$ models. We concluded that the GSL
is satisfied for both models from the early times to the present
epoch. But in the future, the GSL is violated for model 1 when
$T/T_0\in(0,0.133)$ and $(0.569,0.718)$ and for model 2 when
$T/T_0\in(0.005,0.017)$, $(0.039,0.321)$ and $(0.603,0.719)$.
%--------------------------------------------------------------------------------------------------------------------
\subsection*{Acknowledgements}
The authors thank the anonymous referee for a number of valuable
suggestions. The work of K. Karami has been supported financially by
Center for Excellence in Astronomy \& Astrophysics (CEAA-RIAAM),
Maragha, Iran.
%-----------------------------------------------------------------------------------------------

%-----------------------------------------------------------------------------------------------
\clearpage
%-----------------------------------------------------------------------------------------------
%\clearpage
%-----------------------------------------------------------------------------------------------
\clearpage
%-----------------------------------------------------------------------------------------------
%\clearpage
%-----------------------------------------------------------------------------------------------
\clearpage
%-----------------------------------------------------------------------------------------------
%\clearpage
%-----------------------------------------------------------------------------------------------
\clearpage
%-----------------------------------------------------------------------------------------------
%\clearpage
%-----------------------------------------------------------------------------------------------
\clearpage
%-----------------------------------------------------------------------------------------------
%\clearpage
%-----------------------------------------------------------------------------------------------

\end{document}